# Various approaches to a human preference analysis in a digital signage display design

*Jerzy Grobelny, Rafał Michalski*
Institute of Organization and Management (I23),
Faculty of Computer Science and Management (W8),
Wrocław University of Technology,
27 Wybrzeże Wyspiańskiego,
50-370 Wrocław, Poland

## Abstract

The paper is concerned with various ways of analysing the subjective assessment of displaying digital signage content. In the beginning the brief description of the signage system evolution is described and next, the carried out experiment is depicted. The preferences of the 32 subjects were obtained using pairwise comparisons of the designed screen formats. Then the priorities were derived by applying the Analytic Hierarchy Process (Saaty, 1977; 1980) framework. The gathered data were modelled and analysed by means of the analysis of variance, multiple regression, conjoint and factor analysis. The results suggest that the application of different methods of preference analysis may provide additional information, which could facilitate more in-depth understanding of the given preference structure.

# Introduction

The process of human decision making is very often influenced to a large extent by individuals' preferences. Because this phenomenon is observed in many areas, the preferences are subject to investigation by numerous researchers from various fields of science including for instance biology, economy, medicine, psychology, sociology and human factors. The present study is directly focused on the various ways of analysing human preferences towards different display designs used in digital signage systems. The digital signage systems are not only becoming commonplace but also evolve towards more and more interactive solutions. Hence, the scientific examination in this area may be carried out from the perspective of the human-computer interaction (HCI) field of study, where the investigation of people's attitudes and subjective feelings are also one of the major areas of interest.

The next section presents some general information about the evolution of signage systems. Later, a brief description of the ways of the preference evaluation is provided. In the consecutive part of this paper the experimental data related with the digital signage display design preferences are used to conduct analyses by means of various methods. They include the classical analysis of variance, multiple regression as well as conjoint and factor analyses. The discussion and concluding remarks are given in the remainder of the article.

# Evolution of signage systems

Signage systems have been a very popular way of conveying information for hundreds of years. Posters, bills, banners, flags, and many other conventional means of still image communication (Access Displays, 2006) have been extensively used not only in almost any branch of industry or services but also for charity and social projects. Even in the contemporary computerized world, the paper-based graphical message is willingly taken advantage of especially in various widely-understood marketing activities.

During the years of experience in this area a number of innovation that aimed at attracting peoples' attention more effectively were introduced, for instance: popup displays, spinning and swing signs (Warpfive International, 2009; Sign Spin, 2007), roll-up screens (Exact, 2010). Further technical improvements resulted in constructing systems that allow for changing the display content mechanically, for example by scrolling multiple posters (Bei Dou Xing Science & Technology Development, 2010) or by rotating the triangular louvers that enables showing different messages as the three faces are exposed (Triple Sign System AB, 2010). The information technology revolution has recently opened new opportunities for advertisers. For many years, the marketing practitioners and researches focused on the digital solutions based on the internet that are usually tailored to an individual consumer. The technology progress in digital displaying devices along with the networks becoming widespread facilitated the development of electronically based solutions for out-of-home advertising. These digital signage solutions were probably originated by comparatively simple monochromatic led signs that were able to scroll text, present simple graphics and provided greater flexibility. These solutions were later considerably enhanced and nowadays are available in numerous versions (e.g. Signs Plus LEDs, 2010). As the costs of producing LCD or LED panels were continuously dropping down they were gaining more and more interests among advertisers. The process of replacing the conventional bill boards with the digital display solutions is especially intensive in more urbanized areas where the electricity supply and access to computer networks is not a problem and there is a bigger number of potential customers. Thanks to the internet network connections, the individual electronic displays could form together a digital signage network used to present marketing information to consumers at various places in a specific way. Such a solution provides great flexibility in distributing the content to be displayed and allows for the so called advertisement narrowcasting (Harrison & Andrusiewicz, 2004). That narrowcasting enables the advertisers to prepare the appropriate message for the particular segment of customers in given place and time. For example, to display at an airport the content targeted at business travellers on Monday mornings and family-aimed message might appear Friday afternoons. The ongoing technology progress gives also completely new electronically based opportunities to the marketing sector. Some of the modern experimental technologies include mid-air displays (Rakkolainen & Lugmayr, 2007; Rakkolainen, 2008), emergent displays that are blended with their environment (Chandler et al., 2009) or digital equivalent of the cylin-drical sidewalk signboard (Lin et al. 2009).

Easy access to more and more sophisticated and relatively cheap video and computer devices and the pressure for winning the people's attention inclined to incorporate some forms of interaction between digital signage equipment and potential consumers. Thus from one hand the advertisers tend to use the digital signage technologies not only for conveying information but also as a way for eliciting data about potential customers. The gathered data may then allow for the adaptation of the display content to the passers by (Storz et al., 2006; Chen et al. 2009). On the other hand, people more and more often are able to communicate with the signage system using for instance touch sensitive screens, mobile devices (Cheverst et al., 2005) or even hand gestures (Chen et al., 2009). Other interesting forms of interaction may be found at Valli (2010).

Although the digital signage systems are becoming similar to interactive systems, they considerably differ from the standard man-machine solutions especially when the context of use is concerned. As the described domain is relatively new, the design recommendations are based mostly on practitioners' heuristics for example proposed by Bunn, (2009). Some of the design problems concerned with the digital signage were also presented in the documentary film *Helvetica* (2007) directed by Gary Hustwit and focused on typography and graphic design analysed from the perspective of global visual culture. However, there is little scientific systematic research regarding the perception of various kinds of solutions specific for the digital signage systems. In the next section, the preference evaluation process used in this investigation is described.

## Preference evaluation

There is no doubt that in the advertising domain, the customer contentment is considered especially significant, hence this study is devoted to the user's preferences towards various versions of the digital signage display design. Moreover, peoples' preferences are directly connected with the satisfaction, which is one of the main dimensions of assessing the usability of interactive systems in the field of Human–Computer Interaction (ISO 9241, ISO 9126).

It is known from psychology that the preferences may be quite complex and strongly depend on the context. Therefore, taking advantage of various approaches to determine the preference structure seems to be reasonable. There are multiple ways of obtaining the preference data from people and analysing them. The most popular ways of collecting the preferences include direct ranking and pairwise comparisons. In the former method, the user assesses all objects (products, services, alternatives) simultaneously, while the latter one requires the evaluation of two items at a time. In this research the pairwise comparison approach was applied. The technique is usually easier for the user to perform and allows for considerably better accuracy of stimuli estimation (Koczkodaj, 1998). One must remember however, that in this case the number of necessary comparisons grows rapidly with the increasing number of analysed variants. After collecting responses from the subjects, the hierarchy of preferences needs to be derived by means of available procedures. The most popular methods of calculating a priority vector from the numerical pairwise comparison matrix include eigenvalue/eigenvector approach and the logarithmic least squares procedure (Dong et al., 2008). There is no agreement among the researchers which of the techniques is better (compare e.g. Barzilai, 1997, and Saaty & Hu, 1998).

For the purpose of this investigation, the approach advocated by Thomas Saaty (1977, 1980) within the framework of Analytic Hierarchy Process (AHP) was applied. According to this method, the prioritisation of people's relative likings is carried out by finding the principal eigenvector corresponding to the maximal eigenvalue of the symmetric and reciprocal pairwise comparisons matrix. The computed for every person principal eigenvector is normalized in such a way that the sum of all its values equals one. The higher value of relative weight is obtained, the bigger the preference there is for a given alternative. According to the AHP, there is also a possibility of calculating for every subject taking part in the investigation the so called consistency ratio (CR). Higher values of this parameter indicate bigger inconsistencies in pairwise comparisons. Generally, the CR value lower than 0.1 is considered acceptable.

The idea of using the principal eigenvector was employed in the presented research to derive relative subjective preferences towards the analysed designs. The priority vectors were computed individually for every subject and were treated as a main dependent measure. The obtained preferences were next analysed by means of four different techniques, namely the: analysis of variance, multiple regression, conjoint analysis, and factor analysis. In the following paragraphs general information about these methods is demonstrated.

Both the classical analysis of variance as well as the multiple regression with all their modifications are commonly applied by researchers in various areas. The analysis of variance is employed to verify whether the differences in mean values for specific groups are statistically significantly different from each other. In turn, the general purpose of the multiple regression group of methods is to analyze the relationship between a dependent variable and independent predictors. In the context of this study, the derived preferences are treated as a dependent variable whereas the digital signage display design attributes as independent factors.

By means of the conjoint analysis it is possible to decompose the overall preference assessment of a given profile into partial contributions assigned to attributes taken into account during the experiment (Luce & Tukey, 1964; Krantz & Tversky, 1971). Moreover, this method enables the researcher to calculate the part-worths for all attribute levels and the importance of examined attributes. For several decades, the conjoint analysis has been exploited in a variety of research encompassing preference evaluations in many fields of science (Green et al., 2001). Among them marketing and consumer research employed the technique particularly extensive (Green & Srinivasan, 1978; 1990).

The main purpose of the factor analysis is the search for hidden common factors that best account for the covariance structure of the examined variables. The rationale supporting this approach is that in real studies the directly observed factors are quite often influenced by some other effects that are not directly measured. Mulaik (1986) provides the detailed review of the factor analysis evolution from 1940 until the mid eighties. In the work of Steiger (1994) the later developments are presented. In this study the factor analysis allows to get the fuller picture of the participants' preferences, and to examine if they are influenced by the specified attributes differently than it was assumed by the experimental design.

## The experiment

The following subsections include a description of the carried out experiment concerned with the digital signage. First, the methods used are presented, and then the results are analysed.

### Methods

**Subjects**

Nineteen women (59%) and thirteen men (41%) at the average age of 23.3 years and standard deviation of 1.5 year took part in this experiment. The youngest participant was 19 while the oldest 27 years old. The volunteers were from Computer Science and Management Faculty of Wrocław University of Technology (17 students, 53%) and from the Form Design Department of Wrocław Academy of Art and Design (15 participants, 47%). The subjects reported spending from three to 16 hours a day operating the digital devices, with the mean equalled to eight and a half hours and standard deviation 2.1 hrs.

**Experimental design and procedure**

Two independent factors were analysed in the current research: the space between three parts of the screen layout used for informative purposes and the type of the background visible between these areas of data presentation. The first variable was specified at three categorical levels: small, medium, and large. In the second factor, three different backgrounds were used: vivid colours and with a gaudy facture, simple rectangular and colourful shapes, and the uniform colour. The three gap sizes along with the three various background types produced nine variants. The full factorial design of this simple experiment allowed estimating interaction effects between factors. The experimental conditions of all studied digital signage screen profiles are illustrated in figure 1.

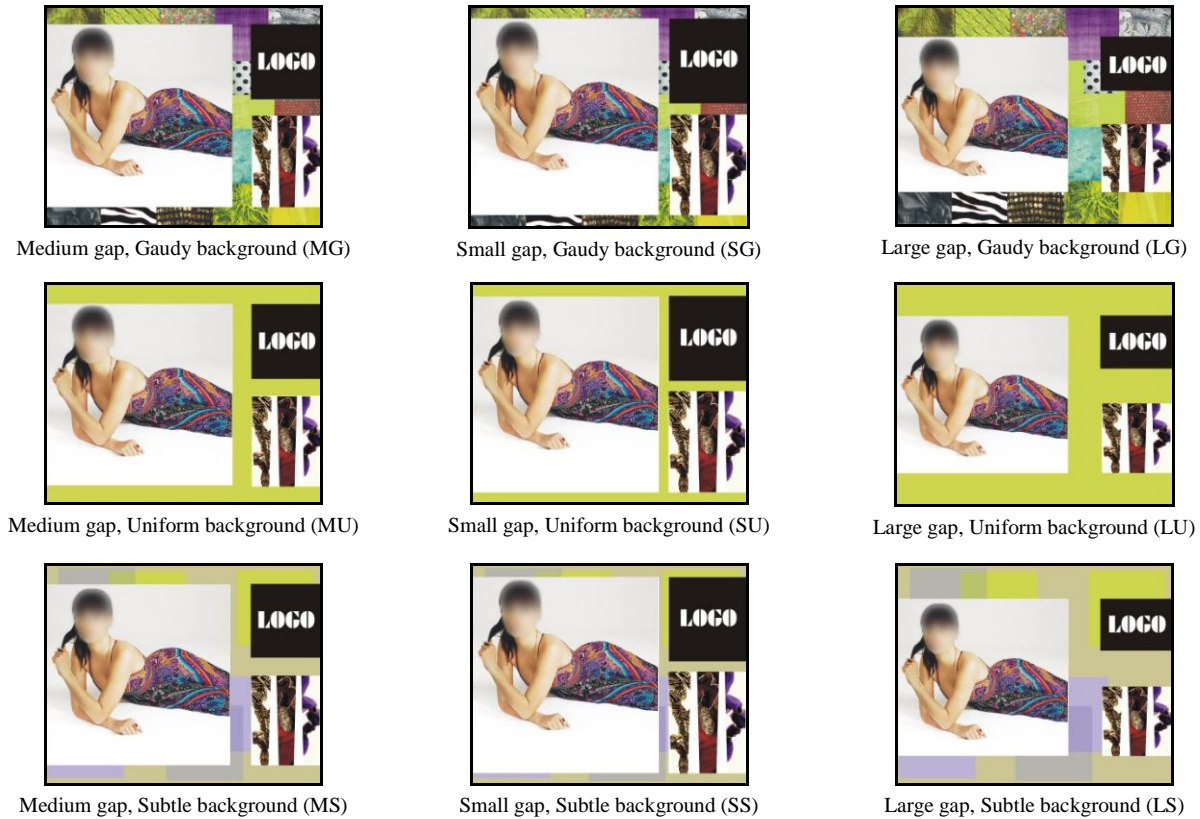

Fig. 1 All nine studied digital signage screen layout profiles.

A standard within subjects model design was employed to investigate the participants' preferences thus all prepared variants of the digital signage screens were evaluated by every participant. Before the study began, the detailed purpose and a scope of the research were explained. After answering personal data questions, the subjects were doing the pairwise comparisons of the screen designs by grading the word *preferred*. All possible combinations of the screen design pairs were demonstrated in random order.

The experiments were conducted in teaching laboratories on similar personal computers equipped with 17" LCD monitors. The resolution was set at 1280 by 1024 pixels. Microsoft® Office PowerPoint® 2003 software was used to display the images of the screen versions, while a paper version of a questionnaire was employed to gather information about the subjects' preferences and to obtain answers to personal questions. The exemplary slide layout along with several rows of the paper questionnaire is given in figure 2.

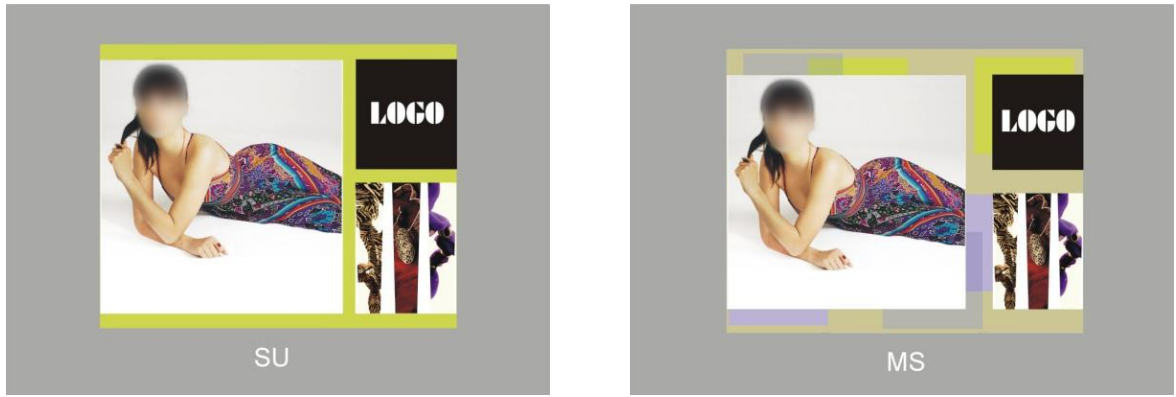

Fig. 2 Exemplary presentation slide (a) and a row from a questionnaire form used in AHP procedure (b).

## Results

**Preference weights and consistency analysis**

The consistency of pairwise comparisons performed during the research was assessed by consistency ratio (CR) values calculated according to the Saaty's procedure. This indicator ranged from 0.031 up to even 0.537. The overall average amounted to 0.173 with the standard deviation at a level of 0.122. The one way Anova showed no differences in mean consistency ratios between subjects from the two universities ($F(1, 30) = 0.4$, $p = 0.53$). Also the age of the participants did not considerably influenced the analysed coherence level ($F(7, 19) = 1.81$, $p = 0.144$). The consistency ratios, however were significantly lower for men then for women ($F(1, 30) = 11.1$, $p = 0.0023$). The average value for males amounted to 0.098 with mean standard error (MSE) 0.016, whereas for females was more than two times higher 0.225 (MSE = 0.029).

In the AHP approach the consistency ratio is required not to exceed the 0.1 value. However, it should be noted that this threshold was recommended by Saaty arbitrarily. Moreover, in the full AHP procedure, if the CR exceeds that limit, the subject is allowed to repeat and correct the preferences, if necessary. In this study procedure this was not the case, so the drop off value of CR was arbitrarily set at the level of 0.2. The application of this criterion, resulted in exclusion of 12 persons from further preference examination, and thus the results of only 20 participants were subject to research in the next analyses. They included 12 men and eight females, eleven persons were from Wrocław University of Technology and nine from Wrocław Academy of Art and Design. The basic descriptive statistics of the obtained relative preferences for all screen profiles are given in table 1.

**Tab. 1** Descriptive statistics of the relative preferences for all screen variants.

| Profile | Gap | Background | Mean | Geometric mean | Standard deviation | Mean standard error | Min | Max | Median |
|---|---|---|---|---|---|---|---|---|---|
| MG | Medium | Gaudy | 0.167 | 0.115 | 0.121 | 0.027 | 0.023 | 0.406 | 0.170 |
| SG | Small | Gaudy | 0.152 | 0.094 | 0.128 | 0.028 | 0.011 | 0.401 | 0.141 |
| LG | Large | Gaudy | 0.132 | 0.090 | 0.104 | 0.023 | 0.014 | 0.375 | 0.111 |
| MU | Medium | Uniform | 0.084 | 0.063 | 0.066 | 0.015 | 0.011 | 0.229 | 0.063 |
| SU | Small | Uniform | 0.099 | 0.069 | 0.097 | 0.022 | 0.015 | 0.363 | 0.066 |
| LU | Large | Uniform | 0.060 | 0.048 | 0.039 | 0.009 | 0.009 | 0.131 | 0.048 |
| MS | Medium | Subtle | 0.097 | 0.071 | 0.074 | 0.016 | 0.018 | 0.238 | 0.053 |
| SS | Small | Subtle | 0.097 | 0.065 | 0.081 | 0.018 | 0.010 | 0.304 | 0.059 |
| LS | Large | Subtle | 0.112 | 0.066 | 0.113 | 0.025 | 0.011 | 0.409 | 0.072 |

According to the calculated arithmetic and geometric mean as well as a median value, the most preferred was the screen version with the medium gap and gaudy background (MG). On the other hand, taking into account the same parameters, the least liked was the variant with large gap and uniform background (LU). The biggest standard deviation was registered for the medium gap and uniform background (MU), whereas the smallest was computed for the large gap and uniform background (LU). The biggest range (0.398) was observed for the screen with large gap and subtle background (LS), while the smallest range (0.122) occurred for the large gap and uniform background (LU) variant.

**Analysis of variance**

Two way analysis of variance was used to verify if the effects of the gap size, background type, and the interaction between these two variables considerably influenced the mean relative preferences. The obtained results are demonstrated in table 2 and revealed that only the background factor was statistically significant ($p = 0.00034$).

**Tab. 2** Two-way Anova results of the relative weights.

| Factor | SS | df | MSS | F | P |
|---|---|---|---|---|---|
| Gap size | 0.0086 | 2 | 0.0043 | 0.47 | 0.62 |
| Background[*] | 0.15 | 2 | 0.076 | 8.4 | 0.00034 |
| Gap size × Background | 0.022 | 4 | 0.0056 | 0.62 | 0.65 |
| Error | 1.5 | 171 | 0.0091 | | |

[*] $p < 0.0005$

The graphical illustration of the mean AHP weights along with the mean standard errors for the statistically significant factor is presented in figure 3.

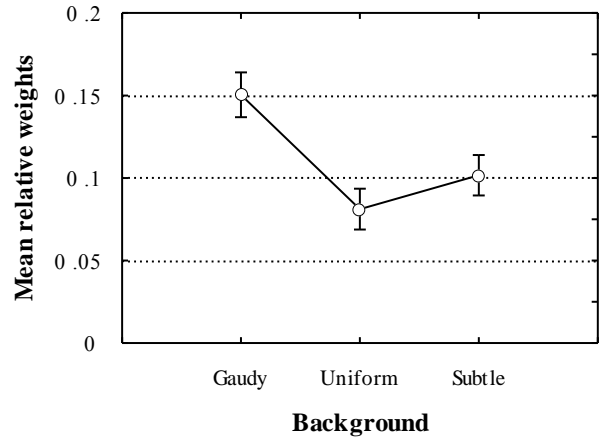

Fig. 3 Mean relative weights depending on screen background type, $F(2, 171) = 8.4$, $p < 0.0005$. Vertical bars denote mean standard errors.

The graph indicates that the gaudy backgrounds were rated the highest, while the uniform type was the least preferred option for the subjects. The difference between the uniform and subtle versions was much smaller (25%) than between gaudy and uniform (85%) or gaudy and subtle one (48%). To check whether these discrepancies are statistically substantial the post-hoc type of analysis was additionally employed. The results are put together in table 3.

**Tab. 3** LSD Post-hoc probabilities for the background effect.

| Factor level | Gaudy | Uniform | Subtle |
|---|---|---|---|
| Gaudy | × | *0.00010 | **0.0057 |
| Uniform |  | × | 0.24 |
| Subtle |  |  | × |

\* $p < 0.0005$
\*\* $p < 0.01$

The LSD post-hoc analysis showed that there was not any statistically significant difference between mean preferences for the profiles with uniform and subtle backgrounds. The tests also revealed that the screen designs with gaudy background were substantially better perceived by the subjects in comparison both with the uniform and subtle backgrounds.

**Multiple regression**

The multiple regression in this investigation was employed to obtain the relationship between the independent variables gap dimension and background type and the dependent measure of the preferences. The two independent factors analysed in the study were categorical, so the artificial coding was applied (Dielman, 2001). In both cases, the values -1, 0, and 1 were applied to represent all levels for the independent effects. The geometric means of preference weights calculated in accordance with the AHP approach served as the dependent variable. The parameters of the regression model were estimated by means of the least square method. The obtained model took the following form:

$$\textit{Geometric Mean Weights} = 0.076 - 0.0041 \cdot \textit{Gap} - 0.020 \cdot \textit{Background}$$

For the presented model, the determinant coefficient amounted to 74%, which means that the variates included in the model explain 74% of the dependent variable variance. The $R^2$ considerably differed from zero $F(2, 6) = 8.7$, $p < 0.017$. The observed preferences along with values calculated from the constructed model, are illustrated in figure 4.

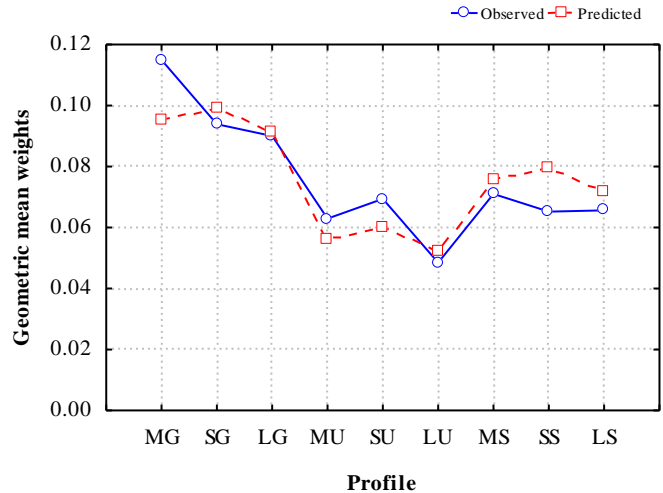

Fig. 4 Predicted and observed geometric means of relative weights for all screen variants.
The regression model with two variables. $R^2 = 74\%$, $F(2, 6) = 8.7$, $p < 0.017$.

A standard procedure of analysing the quality of the proposed model includes the verification whether the parameters differ significantly from zero. The results of this analysis are demonstrated in table 4 and they showed that the Gap variate parameter was statistically insignificant ($p = 0.43$).

Tab. 4 Multiple regression results for relative weights as a dependent variate and Gap and Background as independent variables.

| Variate | Parameter | Std error | t-statistic | p-level |
|---|---|---|---|---|
| Intercept* | 0.076 | 0.0039 | 19.3 | 0.000001 |
| Gap | -0.0041 | 0.0048 | -0.84 | 0.43 |
| Background** | -0.020 | 0.0048 | -4.1 | 0.0064 |

* $p < 0.00001$
** $p < 0.01$

After excluding the Gap factor, the model took the following form:

*Geometric Mean Weights* = 0.076 - 0.020 · *Background*

In this case the R square equalled 71.4%, adjusted 67.3% ($F(1, 7) = 17.5$, $p < 0.005$) and all the parameters were considerably different from zero ($\alpha < 0.005$). The characteristics of the model are given in table 5 and the graphical illustration of the predicted and observed geometric means of relative weights is presented in figure 5.

**Tab. 5** Multiple regression results for relative weights as a dependent variate and Background as an independent variable.

| Variate | Parameter | Std error | t-statistic | p-level |
|---|---|---|---|---|
| Intercept[*] | 0.076 | 0.0038 | 19.7 | 0.000001 |
| Background[**] | -0.020 | 0.0047 | -4.2 | 0.0041 |

[*] $p < 0.00001$
[**] $p < 0.01$

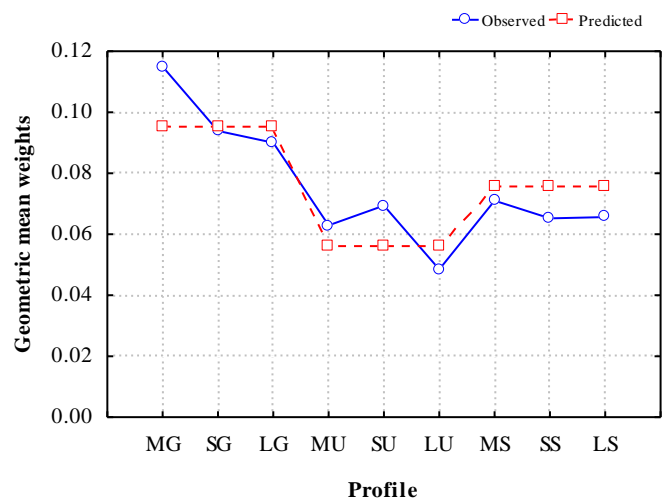

Fig. 5 Predicted and observed geometric means of relative weights for all screen variants.
The regression model with one variable. $R^2 = 71.4\%$, $F(1, 7) = 17.5$, $p < 0.005$.

**Conjoint analysis**

The conjoint analysis in this study was conducted by applying the dummy variable regression for every participant. The AHP relative weights were used as the aggregate response for the individual profile of the digital signage screen format. The individual-level outcomes along with the basic statistical parameters are put together in table 6.

Grobelny J., Michalski R. (2011). Various approaches to a human preference analysis in a digital signage display design, Human Factors and Ergonomics in Manufacturing & Service Industries, 21(6), 529-542. doi: 10.1002/hfm.20295, http://ergonomia.ioz.pwr.wroc.pl/en/**Tab. 6** Estimated partial utilities for individual subject together with basic statistics.

| Variables | 1 | 2 | 3 | 4 | 5 | 6 | 7 | 8 | 9 | 10 | 11 | 12 | 13 | 14 | 15 | 16 | 17 | 18 | 19 | 20 |
|---|---|---|---|---|---|---|---|---|---|---|---|---|---|---|---|---|---|---|---|---|
| **Gap size** | | | | | | | | | | | | | | | | | | | | |
| Medium | 0.043 | 0.031 | -0.047 | -0.029 | 0.007 | 0.012 | -0.015 | 0.031 | -0.014 | -0.013 | -0.020 | 0.049 | 0.030 | -0.008 | 0.010 | -0.029 | 0.037 | -0.025 | 0.008 | 0.037 |
| Small | -0.032 | 0.024 | -0.096 | 0.015 | 0.052 | 0.074 | 0.045 | -0.023 | -0.029 | -0.019 | -0.007 | 0.001 | -0.004 | 0.034 | 0.023 | -0.017 | 0.009 | -0.017 | 0.060 | 0.009 |
| Large | -0.011 | -0.055 | 0.143 | 0.014 | -0.059 | -0.086 | -0.029 | -0.008 | 0.043 | 0.032 | 0.027 | -0.050 | -0.026 | -0.026 | -0.033 | 0.046 | -0.046 | 0.042 | -0.068 | -0.045 |
| **Background** | | | | | | | | | | | | | | | | | | | | |
| Gaudy | 0.108 | 0.059 | -0.006 | -0.064 | -0.090 | 0.121 | 0.144 | 0.027 | 0.077 | 0.018 | 0.145 | 0.146 | 0.022 | -0.078 | 0.146 | 0.132 | 0.123 | -0.075 | -0.090 | -0.077 |
| Uniform | -0.025 | -0.099 | -0.034 | -0.039 | 0.104 | -0.047 | -0.066 | -0.061 | -0.036 | -0.092 | -0.047 | -0.069 | 0.044 | -0.047 | -0.058 | -0.071 | -0.047 | -0.026 | 0.102 | 0.017 |
| Subtle | -0.083 | 0.040 | 0.040 | 0.103 | -0.014 | -0.074 | -0.078 | 0.035 | -0.041 | 0.074 | -0.098 | -0.077 | -0.066 | 0.126 | -0.088 | -0.060 | -0.075 | 0.102 | -0.012 | 0.060 |
| **$R^2$** | 79% | 81% | 73% | 89% | 74% | 77% | 88% | 65% | 92% | 85% | 96% | 77% | 89% | 93% | 97% | 83% | 82% | 84% | 82% | 81% |
| **F** | 3.8 | 4.4 | 2.7 | 8.3 | 2.8 | 3.4 | 7.3 | 1.9 | 11.4 | 5.8 | 21.7 | 3.3 | 7.9 | 13.2 | 35.6 | 4.9 | 4.6 | 5.2 | 4.5 | 4.2 |
| **p** | 0.112 | *0.091 | 0.180 | **0.032 | 0.168 | 0.132 | **0.040 | 0.280 | **0.018 | *0.058 | **0.006 | 0.139 | **0.035 | **0.014 | **0.002 | *0.078 | *0.085 | *0.069 | *0.089 | *0.095 |
| **Gap part-worths** | 28% | 35% | 76% | 21% | 36% | 45% | 25% | 36% | 38% | 24% | 16% | 31% | 33% | 23% | 19% | 27% | 30% | 27% | 40% | 37% |
| **Background part-worths** | 72% | 65% | 24% | 79% | 64% | 55% | 75% | 64% | 62% | 76% | 84% | 69% | 67% | 77% | 81% | 73% | 70% | 73% | 60% | 63% |

*p < 0.1
**p < 0.05

The average relative importances along with the mean part-worths were calculated and the results are demonstrated in table 7.

**Tab. 7** Aggregate-level relative importances and part-worth estimates.

| Variables | Relative importance | Part-worth estimates |
|---|---|---|
| **Gap size** | 32.5% | |
| Medium | | 0.00478 |
| Small | | 0.00501 |
| Large | | -0.00978 |
| **Background type** | 67.5% | |
| Gaudy | | 0.0393 |
| Uniform | | -0.0299 |
| Subtle | | -0.00939 |

The *R*-squared values calculated for every subject's regression served as a goodness-of-fit criterion, and average value of this parameter equalled to 83%. Mean *F* statistics for all of the regressions amounted to 7.84 whereas the average significance level $p = 0.086$.

The above analysis results allow for choosing the most preferred digital signage screen version. There are several possible decision rules that can be applied. Among the most popular are the first choice model (FCM), Bradley, Terry, and Luce (BTL) probability choice model, and logit probability model (LPM). In the first choice model the percentage of all subjects that have rated the given variant the highest serves as the selection criterion. In the Bradley-Terry-Luce model the choice probability for a given person is calculated by dividing the utility of this profile by the sum of utilities of all profiles. Then the individual probabilities are averaged across all subjects. According to the recommendation formulated in the AHP approach, the geometric instead of arithmetic means were employed in this choice simulation method. The logit model estimates the choice probabilities in similar way as the BTL approach but before making the division, the numerator is computed by raising the Euler's constant to the power of the appropriate utility, and the denominator is a sum of the Euler's constants raised to the power of all utilities. In this study all persons for whom any of the negative predicted value equalled zero were excluded from computations both in the BTL and LPM models. The results of the three described simulation models are demonstrated in table 8.

**Tab. 8** The choice simulation results for different models.

| Profile | Gap | Background | FCM | BTL | LPM |
|---|---|---|---|---|---|
| MG | Medium | Gaudy | 20% | 0.0807 | 0.1128 |
| SG | Small | Gaudy | 15% | 0.1033 | 0.1134 |
| LG | Large | Gaudy | 15% | 0.1008 | 0.1148 |
| MU | Medium | Uniform | 5% | 0.0404 | 0.1053 |
| SU | Small | Uniform | 10% | 0.0328 | 0.1059 |
| LU | Large | Uniform | 0% | 0.0641 | 0.1071 |
| MS | Medium | Subtle | 10% | 0.0901 | 0.1126 |
| SS | Small | Subtle | 10% | 0.1010 | 0.1134 |
| LS | Large | Subtle | 15% | 0.0865 | 0.1147 |

Application of the maximum utility criterion as a decision rule would result in selecting the profile with a gaudy background and a medium gap. The BTL (Bradley, Terry, and Luce) probability choice model would recommend the SG profile (small gap with gaudy background) while the logit approach LG screen (large gap and gaudy background). Thus generally, it can be said that the presented models suggest using the screen design with gaudy background, however the choice between gap types is not clear-cut and depend on the choice model.

**Factor analysis**

To analyse the obtained experimental results from a different point of view, the factor analysis as a classification method was employed (Hill & Lewicki, 2007). The relative weights obtained by means of the AHP method for all digital signage screen variants were used as input values. The covariance matrix that shows relations between the design profiles is presented in table 9.

**Tab. 9** Covariance matrix of preference weights computed for all studied profiles that served as an input to the factor analysis.

|    | MG      | SG      | LG      | MU      | SU      | LU      | MS     | SS     | LS     |
|----|---------|---------|---------|---------|---------|---------|--------|--------|--------|
| MG | 0.0145  |         |         |         |         |         |        |        |        |
| SG | 0.0108  | 0.0163  |         |         |         |         |        |        |        |
| LG | 0.0033  | 0.0032  | 0.0108  |         |         |         |        |        |        |
| MU | -0.0033 | -0.0034 | -0.0030 | 0.0043  |         |         |        |        |        |
| SU | -0.0058 | -0.0052 | -0.0058 | 0.0055  | 0.0093  |         |        |        |        |
| LU | -0.0002 | -0.0012 | 0.0006  | 0.0008  | 0.0002  | 0.0015  |        |        |        |
| MS | -0.0052 | -0.0053 | -0.0042 | -0.0001 | 0.0014  | -0.0011 | 0.0055 |        |        |
| SS | -0.0058 | -0.0059 | -0.0043 | -0.0002 | 0.0018  | -0.0011 | 0.0053 | 0.0066 |        |
| LS | -0.0084 | -0.0093 | -0.0006 | -0.0007 | -0.0015 | 0.0006  | 0.0036 | 0.0037 | 0.0127 |

Based on these data, multiple factor analyses with maximum likelihood method of factors' extraction followed by the normalized orthogonal varimax rotation were conducted to find possibly the best factor loading structure. The findings of the analysis of variance and multiple regression showed that the gap factor influenced the subjects' preferences only to a small degree so it was checked whether the covariance matrix structure could be reasonably explained by a single factor related with the type of the background. Such a factor could be called "degree of gaudiness" for instance. The results of this approach are given in the fourth and fifth column of table 10.

**Tab. 10** Factor loadings, communalities and proportion of variance explained obtained by maximum likelihood factor analysis followed by normalized varimax rotation for one, two, and three factors' structure.

| Screen variants | | | One factor structure | | Two factors' structure | | | Three factors' structure | | | |
|---|---|---|---|---|---|---|---|---|---|---|---|
| Label | Gap | Background | F1 | Com. | F1 | F2 | Com. | F1 | F2 | F3 | Com. |
| MG | Medium | Gaudy | *0.654* | 0.428 | -0.559 | 0.515 | 0.578 | -0.396 | 0.195 | *-0.750* | 0.758 |
| SG | Small | Gaudy | *0.630* | 0.397 | -0.539 | 0.492 | 0.533 | -0.357 | 0.131 | *-0.805* | 0.792 |
| LG | Large | Gaudy | **0.551** | 0.304 | **-0.556** | **0.406** | **0.475** | **-0.567** | **0.465** | -0.117 | **0.551** |
| MU | Medium | Uniform | -0.054 | 0.003 | *0.960* | 0.243 | *0.980* | *0.950* | 0.260 | 0.120 | *0.985* |
| SU | Small | Uniform | -0.276 | 0.076 | *0.906* | -0.030 | *0.821* | *0.929* | -0.087 | 0.084 | *0.879* |
| LU | Large | Uniform | 0.326 | 0.106 | 0.220 | **0.408** | 0.215 | 0.140 | **0.591** | 0.205 | **0.412** |
| MS | Medium | Subtle | *-0.942* | *0.888* | 0.215 | *-0.900* | *0.856* | 0.122 | *-0.745* | **0.520** | *0.841* |
| SS | Small | Subtle | *-0.934* | *0.873* | 0.196 | *-0.945* | *0.930* | 0.120 | *-0.827* | **0.502** | *0.950* |
| LS | Large | Subtle | **-0.474** | 0.225 | 0.022 | **-0.470** | 0.221 | -0.217 | 0.019 | *0.907* | *0.870* |
| **Proportion of variance explained:** | | | **36.7%** | | **31.0%** | **31.3%** | | **27.4%** | **21.5%** | **29.3%** | |

Factor loadings in absolute values greater than 0.6 are in italics and bolded, between 0.4 and 0.6 are bolded
Communalities > 0.8 are bolded and in italics, the values between 0.4 and 0.8 are bolded

The interpretation of factor analysis results are usually troublesome, since there is no agreement among researchers as to what value of a factor loading can be treated as high and what is the threshold for suppressing the factor loading. Some investigators suggest 0.3 as the minimum loading of an item (Tabachnick & Fidell, 2001; Hair et al. 1987, 1998). Other researchers classify factor loadings of 0.70 or above as high, the values between 0.51–0.69 as medium and 0.5 or lower are described as low (Kaufman, 1994). One of the most common proposals, applied also in this study, involves treating the absolute value of 0.4 as a cut off, and interpreting the factor loading absolute values of 0.6 as high (Stevens 1986, 1992, 2002; Hair et al. 1998). Apart from the presented rules of thumbs some interesting results were obtained by Peterson (2000). He compared the real factor analysis metadata with randomly generated data, and advised not to use factor loadings below 0.3. Moreover, he also recommends pursuing the solutions in which the variance explained by the factors exceeds 50%. The quality of the factor analysis can additionally be evaluated by analysing the obtained communalities. Velicer and Fava (1998) suggest that values greater than 0.8 for this parameter are considered high. However Costello and Osborne (2005) argue that low to medium communalities between 0.4 and 0.7 are more common in real life data and only variables with communalities lower than 0.4 are not acceptable.

In light of the described recommendations, the factor loadings obtained in this study for the one factor structure are not satisfying. The correlations between the screen variants with a uniform background (MU, SU, LU) do not exceed the 0.4 cut off threshold, and for two others (LG and LS) the medium scores were computed. The proportion of explained variance was merely 37%, and only in three cases (MG, MS, SS) the communalities were above 0.4.

The presented earlier in this paper conjoint analysis outcomes indicate that both of the factors specified in the experimental design are considerably important for the elicited relative preferences. To verify that view, the factor analysis with assumed two factors was applied. However again, the results, which are given in columns 6–8 of table 10, show little support for this standpoint. The high factor loadings were computed only for four variables (MU, SU, MS, SS). Although the proportion of variance explained by the two factors was decent (62.3%), there were multiple and considerable cross-loadings for the screen designs with a gaudy background (MG, SG, LG). Additionally, the levels of communalities left a lot to be desired: two of them were unacceptable (LU, LS) and only four were bigger than 0.8. The results obtained for the three factors' structure demonstrated in the last four columns of the table 10 seem to be of much better quality than the previous two analyses. All the communalities are above the minimal level

Grobelny J., Michalski R. (2011). Various approaches to a human preference analysis in a digital signage display design, Human Factors and Ergonomics in Manufacturing & Service Industries, 21(6), 529-542. doi: 10.1002/hfm.20295, http://ergonomia.ioz.pwr.wroc.pl/en/and the proportion of the common variance explained by the three factors was close to 80%. Although the factor loadings seem to be reasonably high, the significant cross-loadings between the first and second factor for the LG screen variant as well as between the second and third factor for MS and SS profiles, make the findings look ambiguous and difficult to interpret.

The most interesting outcomes were obtained by applying the hierarchical analysis of oblique factors (Thurstone, 1947; Thomson, 1948; Schmid & Leiman, 1957) to the covariance matrix of preferences from table 9. The resulting factor loadings computed according to this approach and demonstrated in table 11, suggest that there are three primary factors that characterise the analysed variables. There also appear to be one additional, and more general secondary factor, which is considerably correlated with the second and third primary factors and much less related with the first one.

**Tab. 11** Secondary and primary factor loadings obtained by applying the hierarchical analysis of oblique factors.

| Label | Gap    | Background | Secondary 1 | Primary 1 | Primary 2 | Primary 3 |
|-------|--------|------------|-------------|-----------|-----------|-----------|
| MG    | Medium | Gaudy      | ***-0.647***    | 0.254     | -0.005    | **-0.524**    |
| SG    | Small  | Gaudy      | ***-0.634***    | 0.216     | -0.066    | **-0.582**    |
| LG    | Large  | Gaudy      | **-0.478**      | **0.465**     | 0.323     | 0.043     |
| MU    | Medium | Uniform    | 0.255       | ***-0.899***  | 0.331     | 0.040     |
| SU    | Small  | Uniform    | 0.396       | ***-0.848***  | 0.026     | -0.044    |
| LU    | Large  | Uniform    | -0.125      | -0.167    | **0.554**     | 0.247     |
| MS    | Medium | Subtle     | ***0.694***     | 0.030     | **-0.531**    | 0.277     |
| SS    | Small  | Subtle     | ***0.723***     | 0.039     | ***-0.604***  | 0.250     |
| LS    | Large  | Subtle     | **0.430**       | 0.316     | 0.160     | ***0.748***   |

Factor loadings in absolute values greater than 0.6 are in italics and bolded, between 0.4 and 0.6 are bolded

The presented structure seems to best represent the set of screen variants analysed in this study, since the primary factor loadings are moderate or high (none is below the 0.4 threshold), and the cross-factor distribution is not meaningful. The proposal, however, is definitely more complex than the structures that could be possibly expected from the experimental design applied in this research.

## Discussion and conclusions

Unquestionably, the people's preferences play a significant role in a decision making process, thus examining, modelling and determining the real structure of them is essential in many areas. Understanding the users' attitudes and finding the best possible ways of analysing them seems to be very important also in the field of usability of interactive systems, especially in its satisfaction dimension.

The main focus of this study was to thoroughly examine users' preferences towards some of the screen characteristics of the digital signage displays using various methods. The analysed graphical solutions were differentiated by two factors: the background type and the amount of the free space between different visual components of the screen layout. For retrieving the relative preferences, the Saaty's AHP framework was applied. The obtained findings show that depending on the approach, the investigator may come to various conclusions and make different practical decisions.

The results of Anova analysis proved that neither the gap factor nor the interaction between the gap and background type had a substantial influence on the obtained mean preferences. The further post-hoc analysis additionally demonstrated that there was not any meaningful difference between the uniform and subtle backgrounds. These find-Grobelny J., Michalski R. (2011). Various approaches to a human preference analysis in a digital signage display design, Human Factors and Ergonomics in Manufacturing & Service Industries, 21(6), 529-542. doi: 10.1002/hfm.20295, http://ergonomia.ioz.pwr.wroc.pl/en/ings would recommend the researcher to choose any of the screen layouts with a gaudy background. Similar suggestion was obtained in the multiple regression approach, where the AHP preference weights were presented as a function of the independent variables. The initial two variables model proved to be inadequate due to the insignificance of the Gap parameter. Therefore, the final formula included only the background as a dependent variate. This result supported the view of taking into consideration mainly the background type while neglecting the gap size factor in the design of digital signage graphical layouts. The findings yielded from the conjoint analysis are only partly in concordance with the previous ones. They still show that the background variable is the most significant with the relative importance equalled 67.5%, but the relative importance of 32.5% acquired for the gap size indicates that this factor could have a significant impact on the users' preferences as well. The series of conducted factor analyses provided some evidence that the preference structure in this research could be much more complicated than it was observed by means of the three techniques described earlier on. In particular, the hierarchical analysis of the preference weights covariance matrix resulted in as many as three primary latent factors and one, more general, secondary factor. These findings indicate, that the preferences concerned with the examined factors may by interrelated or may be influenced by some other factors that were not controlled in this study. This interpretation could have been omitted relying solely on the outcomes of the regression analysis, where the interaction between the examined factors was statistically insignificant. The brief summary of the findings along with the conclusions resulting from the applied in this study individual methods are put together in table 12.

**Tab. 12** Summary of the results from all of the employed methods.

| Method | Results | Conclusions |
|---|---|---|
| Analysis of variance | • Background: significantly important ($p < 0.0005$)<br>• Gap and the Gap-Background interaction: statistically insignificant ($\alpha = 0.05$)<br>• Post-hoc analysis for Background:<br>  – Gaudy significantly better perceived than Uniform ($p < 0.0005$) and Subtle ($p < 0.01$)<br>  – The difference between Uniform and Subtle was not meaningful ($\alpha = 0.05$) | • Not take into consideration the Gap effect<br>• Focus on the Gaudy background, which is the best, and do not differentiate between the Uniform and Subtle backgrounds. |
| Multiple regression | • The regression formula after excluding the Gap variate because of the its statistically insignificant parameter ($\alpha = 0.05$):<br>Geometric Mean Weights = 0.076 - 0.020 • Background<br>• $R^2 = 71.4\%$, $F(1, 7) = 17.5$, $p < 0.005$ | • Not take into account the Gap effect<br>• Gaudy backgrounds are better perceived than Subtle ones, and Subtle backgrounds are higher rated than Uniform ones. |
| Conjoint analysis | • Relative importances (RI) and part-worth estimates (PW):<br>  – Gap: RI = 32.5%; PWMedium = 0.00478, PWSmall = 0.00501, PWLarge = -0.00978<br>  – Background: RI = 67.5%; PWGaudy = 0.0393, PWUniform = -0.0299, PWSubtle = -0.00939<br>• Best variants according to choice simulation models:<br>  – FCM: medium gap with gaudy background (MG)<br>  – BTL: small gap with gaudy background SG<br>  – LPM: large gap with gaudy background (LG) | • The Background variable is far more important than the Gap one, but the Gap should also be included during making practical decisions.<br>• According to choice simulators one should choose the layout with gaudy background. The simulators, however, are inconsistent when the Gap is concerned and provide different recommendations. |
| Factor analysis | • Series of various factor analyses provided no clear structure of the factor loadings. The best was the three factors' structure, however because of the significant cross-loadings the outcome was not acceptable. | • The results indicate that this method might probably be inappropriate for the gathered data. |
| Hierarchical factor analysis | • Suggests three primary factors and one more general secondary factor, which is considerably correlated with the second and third primary factors and much less related with the first one. | • The application of this method revealed of the quite clear general structure of the preferences, which is however different from the design of the experiment. It suggests that the perception of the examined layouts depend on different aspects of the analysed layouts than it was initially assumed in this experiment. |

Generally, in light of the conducted analyses, both of the investigated effects had significant impact on the users' perception of the examined screen layouts. The results also indicate that the background was far more important than the gap variable, however from the hierarchical factor analysis it can be seen that they might be subject to some more general secondary factor.

There are of course many limitations related with this research. First, the inclusion of other factors related with the digital signage screen design could broaden the analysis, which could be of great value especially to practitioners. Also adding more levels to the investigated factors could be interesting. Because of the not fully clear preference structure obtained in this research, the future investigations might include for example, the examination of the optimal geometrical properties and interrelations between the graphical components of digital signage displays. Possibly, finding the existence of the so called golden sections between some of them (Gielo-Perczak, 2001) could result in elaborating the appropriate design recommendations.

The present study was carried out on a comparatively small sample and the reader should be very cautious in generalizing the obtained results, all the more the considerable inconsistencies that were observed particularly in females, additionally decreased the number of subjects used for further examination. Furthermore, in the conjoint analysis, the regression models obtained for some individuals were statistically not significant and some others were significant merely at the level of 0.1. Taking into account the small number of participants in this research they were not excluded, but it may be argued whether or not this was justifiable. The application of the cluster analysis to the conjoint data may be worth noting provided that a bigger sample is available. In applying the factor analysis, it is also possible to use other than in this study rotational approaches or threshold values, which could lead to different preference structure proposals. Moreover, the conducted analyses can be supplemented by employing for instance some version of the path models originally proposed by Wright in 1921.

Despite these limitations, the study undoubtedly shows that even for quite straightforward experimental set-up, the structure and interpretation of users' preferences may be problematic. It is naturally hard to recommend one best approach since each of the used in this study techniques has its advantages and limitations. Therefore, it may be reasonable to apply, various methods of subjective data analysis whenever possible to get a fuller picture of the preference structure. They should rather be used as complementary instead of as mutually exclusive. Such a comprehensive approach, in turn, may help to make correct practical decisions and facilitate to elaborate adequate future research objectives. Furthermore, it seems that the presented discussion and conclusions are not only confined to the digital signage and could have practical implications in other areas, especially concerned with the visual communication.

# References

Barzilai, J. (1997). Deriving weights from pairwise comparison matrices, Journal of the Operational Research Society 48, 1226–1232.

Bunn, L. (2009). Content Creation - Simple Guidelines. http://lylebunn.com/, retrieved on 28 of September 2010.

Cattell, R.B. (1966). The scree test for the number of factors. Multivariate Behavioral Research, 1, 245–276.

Chandler, A., Finney, J., Lewis, C., & Dix A. (2009). Toward emergent technology for blended public displays. UbiComp 2009, Sep 30 – Oct 3, Orlando, Florida, USA.

Chen, Q., Malric, F., Zhang, Y., Abid, M., Cordeiro, A., Petriu, E. M., & Georganas, N. D. (2009). Interacting with digital signage using hand gestures. Proceedings of the 6[th] International Conference on Image Analysis and Recognition (pp. 347–358). Halifax, Nova Scotia, Canada: Springer-Verlag.

Cheverst, K., Dix, A., Fitton, D., Kray, C., Rouncefield, M., Sas, C., Saslis-Lagoudakis, G., et al. (2005). Exploring bluetooth based mobile phone interaction with the hermes photo display. Proceedings of the 7[th] international conference on human–computer interaction with mobile devices & services (pp. 47–54). Salzburg, Austria: ACM.

Costello, A.B., & Osborne, J.W. (2005). Best practices in exploratory factor analysis: four recommendations for getting the most from your analysis. Practical Assessment, Research & Evaluation, 10(7), 1–9.

Dielman, T.E. (2001). Applied regression analysis for business and economics. Pacific Grove: Duxbury.

Dong, Y.C., Xu, Y.F., Li, H.Y., & Dai, M. (2008). A comparative study of the numerical scales and the prioritization methods in AHP. European Journal of Operational Research, 186(1), 229–242.

Gielo-Perczak, K. (2001). The golden section as a harmonizing feature of human dimensions and workplace design. *Theoretical Issues in Ergonomics Science*, 2(4), 336.

Green, P.E., & Srinivasan, V. (1978). Conjoint analysis in consumer research: Issues and outlook. Journal of Consumer Research, 103–123.

Green, P.E., & Srinivasan, V. (1990). Conjoint analysis in marketing: new developments with implications for research and practice. Journal of Marketing, October, 3–19.

Green, P.E., Krieger, A.M., & Wind, Y. (2001). Thirty years of conjoint analysis: Reflections and prospects. Interfaces, 31(2), 56–73.

Harrison, J.V. & Andrusiewicz, A. (2004). A virtual marketplace for advertising narrowcast over digital signage networks. Electronic Commerce Research and Applications, 3(2), 163–175.

Hair, J.F., Anderson, R.E., & Tatham, R.L. (1987). Multivariate data analysis. Ed. 2. MacMillan, New York.

Hair, J.F., Anderson, R.E., Tatham, R.L. & Black, W.C. (1998). Multivariate data analysis with readings. 5th ed.. Englewood Cliffs, NJ: Prentice-Hall.

Hill, T. & Lewicki, P. (2007). STATISTICS Methods and Applications. StatSoft, Tulsa, OK.

Valli, A. (2010), Natural Interaction, http://naturalinteraction.org, retrieved on 12 of October 2010.

Sign Spin (2007), Sign Spin family of products, http://signspin.com/, retrieved on 12 of October 2010.

Access Displays (2006), Access Displays, http://www.accessdisplays.co.uk, retrieved on 12 of October 2010.

Bei Dou Xing Science & Technology Development (2010), BDX Signs, http://www.bdxsigns.com/en/products.asp, retrieved on 12 of October 2010.

Exact (2010), Display Systems, http://www.exact.net.pl/gb/index_gb.html, retrieved on 12 of October 2010.

Signs Plus LEDs (2010), Signs Plus LED Display Signs, http://www.signsplusleds.com, retrieved on 12 of October 2010.

Triple Sign System AB (2010), Triplesign Trivision - Motions engage the attention, http://www.triplesign.com, retrieved on 12 of October 2010.

Warpfive International (2009), Warpfive International - Manufacturers of Point of Sale and Display Systems, http://www.warpfive.eu, retrieved on 12 of October 2010.

Hustwit, G., (2007), Helvetica, A documentary film produced by Swiss Dots, in association with Veer, United Kingdom.

ISO 9126–1 (1998). Software product quality, Part 1: Quality model, International Standard.

ISO 9241–11 (1998). Ergonomic requirements for office work with visual display terminals (VDTs), Part 11: Guidance on usability, International Standard.

Kaufman, A.S. (1994). Intelligent testing with the WISC-III. New York: John Wiley & Sons, Inc.

Koczkodaj, W. (1998). Testing the accuracy enhancement of pairwise comparisons by a Monte Carlo experiment. Journal of Statistical Planning and Inference, 69(1), 21–31.

Krantz, D.H. & Tversky, A. (1971). Conjoint measurement analysis of composition rules in psychology. Psychological Review, 78, 151–169.

Lin, J.-Y., Chen, Y.-Y., Ko, J.-C., Kao, H.S., Chen, W.-H., Tsai, T.-H., Hsu, S.-C., Hung, Y.-P. (2009). i-m-Tube: an Interactive Multi-Resolution Tubular Display, MM'09, October 19–24, Beijing, China.

Luce, D.R. & Tukey, J.W. (1964). Simultaneous conjoint measurement: a new type of fundamental measurement. Journal of Mathematical Psychology, 1, 1–27.

Mulaik, S.A. (1986). Factor analysis and Psychometrika: Major developments. Psychometrika, 51(1), 23–33.

Peterson, R.A. (2000). A meta-analysis of variance accounted for and factor loadings in exploratory factor analysis. Marketing Letters, 11(3), 261–275.

Rakkolainen, I. (2008). Mid-air displays enabling novel user interfaces, SAME'08. October 31, Vancouver, British Columbia, Canada, pp. 25–30.

Rakkolainen, I., Lugmayr, A. (2007). Immaterial display for interactive advertisements. ACE'07, June 13–15, Salzburg, Austria.

Saaty, T.L. (1977). A scaling method for priorities in hierarchical structures. Journal of Mathematical Psychology,

15, 234–281.

Saaty, T.L. (1980). The analytic hierarchy process. New York: McGraw-Hill.

Saaty, T.L., Hu, G. (1998). Ranking by the eigenvector versus other methods in the analytic hierarchy process. Applied Mathematical Letters, 11(4), 121–125.

Schmid, J., & Leiman, J. M. (1957). The development of hierarchical factor solutions. Psychometrika, 22, 53–61.

Steiger, J.H. (1994). Factor Analysis in the 1980's and the 1990's: Some old debates and some new developments. In Borg, I., & Mohler, Peter Ph. (Eds.) Trends and perspectives in empirical social research. Berlin: DeGruyter.

Stevens, J.P., (1986). Applied multivariate statistics for the social sciences. Lawrence Erlbaum, Hillsdale, NJ.

Stevens, J.P., (1992). Applied Multivariate Statistics for the Social Sciences. 2nd Edn., Lawrence Erlbaum, Hillsdale, NJ.

Stevens, J.P. (2002). Applied Multivariate Statistics for the Social Sciences (4th Edition). Mahwah, NJ: Lawrence Erlbaum Associates.

Storz, O., Friday, A., & Davies, N. (2006). Supporting content scheduling on situated public displays. Computers & Graphics, 30(5), 681–691.

Tabachnick, B.G., & Fidell, L.S. (2001). Using Multivariate Statistics. Boston: Allyn and Bacon.

Thomson, G.H., (1948). The factorial analysis of human ability. New York: Houghton, Mifflin.

Thurstone, L.L., (1947). Multiple-factor analysis. Chicago: Univ. Chicago Press.

Velicer, W.F., & Fava, J.L. (1998). Effects of variable and subject sampling on factor pattern recovery. Psychological Methods, 3(2), 231–251.

Wright, S. (1921). Correlation and causation. Journal of Agricultural Research, 20, 557–585.